# Unless Connected to Relativity, the First and Second Laws of Thermodynamics are Incompatible

Jean-Louis Tane (tanejl@aol.com). February 2010
Formerly with the Department of Geology, University Joseph Fourier, Grenoble, France

**Abstract**: The first part of this paper is a condensed synthesis of the matter presented in several previous ones. It begins with an argumentation showing that the first and second laws of thermodynamics are incompatible with one another if they are not connected to relativity. The solution proposed consists of inserting the Einstein mass-energy relation into a general equation that associates both laws.
    The second part deals with some consequences of this new insight and its possible link with gravitation. Despite a slight modification of the usual reasoning, the suggested hypothesis leads to a simplification and extension of the thermodynamic theory and to the idea that relativity is omnipresent around us.

## - 1 - Incompatibility of the first and the second laws in conventional thermodynamics

### 1.1 Irreversibility and reversibility in the case of a work exchange

Let us consider a system defined a gas enclosed in a cylinder fitted with a frictionless piston. If this system is concerned by a mechanical work exchange with its surroundings, the equation describing the general case of an irreversible process is:

$$dW_{irr} = - P_e \, dV \tag{1}$$

where $dV$ represents an elementary volume change, $P_e$ the external pressure and $dW$ the corresponding change in work.

In the case of a reversible process, eq. 1 becomes:

$$dW_{rev} = - P_i \, dV \tag{2}$$

where $P_i$ represents the internal pressure.

Therefore, for a given value of $dV$, the difference $dW_{irr} - dW_{rev}$ can be written through the relation:

$$dW_{irr} = dW_{rev} + dV(P_i - P_e) \tag{3}$$

Since $dV$ is positive when $P_i > P_e$ and negative when $P_i < P_e$, the term $dV(P_e - P_i)$ is always positive, and we get in all conditions the relation:

$$dW_{irr} > dW_{rev} \tag{4}$$

Keeping in mind this information, let us imagine an isolated system made of two gaseous parts designated 1 and 2, separated by a diathermic piston. If the initial pressures $P_1$ and $P_2$ are different, the piston will move until they become equal. Applying eq. 1 to both parts, we get:

$$dW_{irr1} = - P_2 \, dV_1 \tag{5}$$

$$dW_{irr2} = -P_1\, dV_2 \tag{6}$$

Since $dV_2 = -dV_1$, the value $dW_{irrSyst}$ of the whole system is:

$$dW_{irrSyst} = dV_1(P_1 - P_2) \tag{7}$$

Knowing that $dV_1$ is positive when $P_1 > P_2$ and negative when $P_1 < P_2$, we have in all cases:

$$dW_{irrSyst} > 0 \tag{8}$$

In the conventional interpretation of the first law, it is admitted as a postulate that the internal energy U of an isolated system cannot vary, and therefore implies the relation:

$$dU_{irr,Syst} = 0 \tag{9}$$

To reconcile this result with the one given by eq. 8, the only possible solution is in admitting that the positive value of $dW_{irrSyst}$ is compensated by a negative value of another energetic term. Referring the well known formula:

$$dU = dQ + dW \tag{10}$$

we are spontaneously tempted to think about a heat exchange and imagine that it must obeys the condition:

$$dQ_{irrSyst} < 0 \tag{11}$$

and more precisely:

$$dQ_{irrSys} = -dW_{irrSys} \tag{12}$$

in order that we can get:

$$dU_{irrSys} = dQ_{irrSys} + dW_{irrSys} = 0 \tag{13}$$

The existence of a heat exchange within the system can be explained by the fact that the temperature tends to increase in the compressed part and to decrease in the expanded one. In such a case, the heat exchange (through the diathermic piston) is the natural response of the system to restore the equalization of the temperatures.

Is it really possible that the heat exchanges occurring within the system are characterized by the condition $dQ_{irrSyst} < 0$? This important question is discussed below.

### 1.2 Irreversibility and reversibility in the case of a heat exchange

The subject is closely related to the state function S, called entropy, and to the second law of thermodynamics, i.e to the expression:

$$dS = dQ/T + dS_i \tag{14}$$

whose precise meaning is:

$$dS = dQ_{rev}/T_e + dS_i \tag{15}$$

Eq. 15 being an entropy equation, the corresponding energy equation takes the form:

$$T_e dS = dQ + T_e dS_i \tag{16}$$

whose meaning is:

$$dQ_{irr} = dQ_{rev} + T_e dS_i \tag{17}$$

In eq. 17, the term $dS_i$ is known to be positive (fundamental information linked to the second law) and the term $T_e$ too (absolute temperature). Therefore the term $T_e dS_i$ is itself positive, so that we have necessarily:

$$dQ_{irr} > dQ_{rev} \tag{18}$$

This last formula can equally be written:

$$dQ_{irr} = dQ_{rev} + dQ_{add} \tag{19}$$

where $dQ_{add}$ means $dQ_{additional}$ and has a positive value.

Applying eq. 19 to part 1 and part 2 successively, leads to:

$$dQ_{irr1} = dQ_{rev1} + dQ_{add1} \tag{20}$$

$$dQ_{irr2} = dQ_{rev2} + dQ_{add2} \tag{21}$$

where both terms $dQ_{add1}$ and $dQ_{add2}$ are positive

By addition, the value $dQ_{irrSyst}$ of the whole system is:

$$dQ_{irrSyst} = dQ_{revSyst} + dQ_{addSyst} \tag{22}$$

In eq. 22, we have $dQ_{revSyst} = 0$ (because the condition of reversibility implies the equality $dQ_{rev2} = -dQ_{rev1}$).

Observing that $dQ_{addSyst}$ is positive (being defined as $dQ_{addSyst} = dQ_{add1} + dQ_{add2}$), the resulting conclusion is:

$$dQ_{irrSyst} > dQ_{revSyst} \tag{23}$$

that implies itself:

$$dQ_{irrSyst} > 0 \tag{24}$$

This result being in disagreement with the expected one (cf. the last two lines of section 1.1), we are led to the conclusion that something is wrong is the basis of the discussion and needs to be revised.

## 1.3 Irreversibility and reversibility in the general case of an energy exchange

If the conclusion just obtained is recognized as valid, it seems that the only possible solution of the problem is in admitting that, contrary to the postulate in use, the correct formulation of the first law of thermodynamics is not

$$dU_{irrSyst} = dU_{revSyst} \tag{24}$$

but:

$$dU_{irrSyst} = dU_{revSyst} + dU_{addSyst} \tag{25}$$

In eq. 25, the term $dU_{addSyst}$ has a positive value when the system is concerned by internal energy exchanges (irreversibility) and a zero value if it is not the case (reversibility). Knowing that real processes always contain a part of irreversibility, the practical significance of eq. 25 is

$$dU_{irrSyst} > dU_{revSyst} \tag{26}$$

Of course, the insertion of the term $dU_{addSyst}$ in the theory raises the question of the origin of this additional energy. The answer suggested in previous papers ([1], [2]) refers to relativity. According to the Einstein mass-energy relation $E = mc^2$, it can be imagined that the energy created is linked to a correlative disintegration of mass, giving to $dU_{addSyst}$ the significance:

$$dU_{addSyst} = - c^2 dm \tag{27}$$

and to eq. 25 the significance:

$$dU_{irr} = dU_{rev} - c^2 dm \tag{28}$$

In eq. 27 and 28, the minus sign placed in front of the term $c^2 dm$ appears as a necessary condition to give $dU_{addSyst}$ a positive value, in the same manner as a minus sign is inserted in eq. 1 to give dW a positive value.

Among the immediate implications of this new conception is the fact that the terms $dQ_{irr}$ and $dQ_{rev}$ can be defined by equations similar to those used for $dW_{irr}$ and $dW_{rev}$ (eq. 1 and 2). This leads to write as introductive definitions the relations:

$$dQ_{irr} = T_e dS \tag{29}$$

$$dQ_{rev} = T_i dS \tag{30}$$

Since S is a state function, dS has the same value whatever is the level of irreversibility of the heating process, so that the difference $dQ_{irr} - dQ_{rev}$ can be written:

$$dQ_{irr} - dQ_{rev} = dS(T_e - T_i) \tag{31}$$

The terms $T_e$ and $T_i$ being positive (absolute temperatures), eq. 29 and 30 imply that the sign of dS is always that of dQ (which is evidently the same for $dQ_{irr}$ and $dQ_{rev}$).

Having $dQ > 0$ when $T_e > T_i$ and $dQ < 0$ when $T_e < T_i$, the same is true for dS. Therefore, the term $dS(T_e - T_i)$ is always positive and implies the relation:

$$dQ_{irr} > dQ_{rev} \qquad (32)$$

This result being identical to the one already obtained with. 23, it gives an indirect confirmation of the validity of eq. 29 and 30.

- 2 - **The omnipresence of relativity**

**2.1. Preliminary remarks**

It is often admitted that the need of relativity is restricted to processes implying very high speeds. Taking into account the considerations examined above, it appears on the contrary that relativity plays a fundamental role in the thermodynamic theory. Combining this data with the wide usefulness of the laws of thermodynamics, we are led to the conclusion that relativity is omnipresent and can never be neglected.

The important point to keep in mind is that eq. 28 covers both the first and the second laws. The first law, usually understood as meaning $dU_{irr} = dU_{rev}$, is understood here as meaning $dU_{irr} > dU_{rev}$ (eq.26). Correlatively, the second law whose classical transcription is the entropy equation $dS = dQ/T_e + dS_i$, takes now the form of the energy equation $T_e dS = dQ + T_e dS_i$ (eq.16) whose precise meaning is $dU_{irr} = dU_{rev} - c^2 dm$ (eq.28).

As can be seen through the references quoted below, the existence of a link between thermodynamics and relativity has been suggested for a long time ([3], [4]) and remains an actively studied subject ([5], [6], [7], [8], [9]). The originality of the argumentation presented above is probably its simplicity, with the advantage of being accessible to a large scientific readership, not necessarily highly specialized in physics and chemistry. The matter that will be discussed now is pursued in the same perspective. It deals with some possible consequences of the suggested hypothesis in the fields of physico-chemistry, astronomy and biology.

**2.1. Possible consequences in physico-chemistry**

The aim of this section is to show that a simple and general relation can be proposed between the term $dU_{addSyst}$ and the differential $dG$ of the thermodynamic function G (Free Energy). This relation is:

$$dG = - dU_{addSyst} = + c^2 dm \qquad (33)$$

For an easier derivation of eq. 33, the discussion is divided into two steps.

**A) First step**

Let us come back to the gaseous system considered in the first lines of section 1.1. If its volume varies from an initial state $V_1$ to a final state $V_2$, the corresponding work exchange obeys the following peculiarities:

If the process is irreversible (practical case) we have to integrate eq.1 and we get:

$$\Delta W_{irr} = \int_{V_1}^{V_2} - P_e \, dV \tag{34}$$

If $P_e$ is constant, eq. 34 becomes:

$$\Delta W_{irr} = - P_e [\Delta V]_{V_1}^{V_2} \tag{35}$$

If $P_e$ is not constant, it can be written:

$$\Delta W_{irr} = - P_e^* [\Delta V]_{V_1}^{V_2} \tag{36}$$

where $P_e^*$ is the average value of $P_e$ during the process.

For a given process, the term $P_e^*$ represents a mathematical constant. Therefore, even if its value is not known, we have necessarily:

$$dP_e^* = 0 \tag{37}$$

If the process is reversible (limited theoretical case), the same reasoning leads to the conclusion:

$$dP_i^* = 0 \tag{38}$$

In a similar way, the integration of eq. 29 leads to:

$$\Delta Q_{irr} = \int_{S_1}^{S_2} T_e \, dS \tag{39}$$

If $T_e$ is constant, eq. 39 becomes:

$$\Delta Q_{irr} = T_e [\Delta S]_{S_1}^{S_2} \tag{40}$$

If $T_e$ is not constant, it can be written:

$$\Delta Q_{irr} = T_e^* [\Delta S]_{S_1}^{S_2} \tag{41}$$

where $T_e^*$ is the average value of $T_e$ during the process, and implies the condition:

$$dT_e^* = 0 \tag{42}$$

The same situation is true for $T_i$ and leads to the relation:

$$dT_i^* = 0 \tag{43}$$

As a preliminary result of the discussion, the terms $dU_{irr}$, $dU_{rev}$ and $dU_{add}$ corresponding to a thermomechanical process can be written under the forms:

$$dU_{irr} = dW_{irr} + dQ_{irr} \tag{44}$$

$$dU_{rev} = dW_{rev} + dQ_{rev} \tag{45}$$

$$dU_{add} = dU_{irr} - dU_{rev} = -c^2 dm \tag{46}$$

Then taking into account eq. 1, 2, 29 and 30 (and respecting the fact that the expression $dU = TdS - PdV$ is more in use than $dU = -PdV + TdS$), we see that another possible formulation of the triplet just evoked is:

$$dU_{irr} = T_e^* dS - P_e^* dV \tag{47}$$

$$dU_{rev} = T_i^* dS - P_i^* dV \tag{48}$$

$$dU_{add} = (T_e^* dS - P_e^* dV) - (T_i^* dS - P_i^* dV) = -c^2 dm \tag{49}$$

Therefore, an alternative writing of $dU_{rev}$ is:

$$dU_{rev} = (T_e^* dS - P_e^* dV) + c^2 dm \tag{50}$$

**B) Second step**

The function free energy G, is defined by the relation:

$$G = H - TS \tag{51}$$

where:

$$H = U + PV \tag{52}$$

Consequently, the expression of dG is given by the well-known relation:

$$dG = dU + PdV + VdP - TdS - SdT \tag{53}$$

whose meaning is:

$$dG = dU_{rev} + P_e dV + VdP_e - T_e dS - SdT_e \tag{54}$$

Taking into account the considerations already discussed (first step), another possible transcription of eq. 54 is:

$$dG = dU_{rev} + P_e^* dV + VdP_e^* - T_e^* dS - SdT_e^* \tag{55}$$

Now, entering in eq. 55 the value $dU_{rev}$ given by eq. 50, we obtain:

$$dG = (T_e^* dS - P_e^* dV) + c^2 dm + P_e^* dV + V dP_e^* - T_e^* dS - S dT_e^* \qquad (56)$$

After simplification and knowing (from eq. 37 and 42) that $dP_e^*$ and $dT_e^*$ are zero, we are led to:

$$dG = + c^2 dm \qquad (57)$$

whose detailed meaning can also be written as:

$$dU_{addSyst} = - dG = - c^2 dm \qquad (58)$$

It is a fundamental point of thermodynamics that a negative value of dG is the condition of evolution of a system. Presented under the form of eq. 58, this information shows more clearly that the negative value of dG is the sign that an additional energy has been created, which is directly related to a negative value of dm, i.e. to a disintegration of matter.

From the theoretical point of view, this information is of great interest. From the practical point of view, that is to calculate dG, the value $dU_{rev}$ that needs to be inserted in eq. 55 is not the one given by eq. 50, as already done, but the one given by eq. 48. In such a case, eq. 55 takes the form:

$$dG = T_i^* dS - P_i^* dV + P_e^* dV + V dP_e^* - T_e^* dS - S dT_e^* \qquad (59)$$

Then taking into account that $dP_e^*$ and $dT_e^*$ are zero, eq. 59 reduces to:

$$dG = dS (T_i^* - T_e^*) + dV (P_e^* - P_i^*) \qquad (60)$$

whose integrated form is:

$$\Delta G = \Delta S (T_i^* - T_e^*) + \Delta V (P_e^* - P_i^*) \qquad (61)$$

Some elementary examples of the use of eq. 61 have been given in previous papers ([1], [2]).

### 2.1. Possible consequences in astronomy

As noted above, eq. 28, whose expression is $dU_{irr} = dU_{rev} - c^2 dm$, appears as a general formula covering the first and second laws of thermodynamics, thanks to their connection with the Einstein mass-energy relation. The important point of the discussion presented below is that the physico-chemical processes occurring within a system imply a decrease of its mass. One of the effects of the geological events, for example, is a decrease of the mass of the Earth.

When a system evolves from an initial state 1 to a final state 2, an important question is the nature of the symptoms that can be an indicator of the level of irreversibility of the process.

We easily conceive that, for a given process, the decrease in mass is more important in conditions highly irreversibible than in conditions slightly irreversible. The problem is that, in an experimental context, the change in mass is so tiny that it cannot be detected. It is therefore impossible for its own variations to be measured.

A symptom more easily observable is the duration of the process, that must be all the more restricted that the level of irreversibility is higher, all other conditions remaining the same. From this point of view, it seems not excluded that the measure of the duration could give information about the additional energy created and the correlative decrease in mass.

Another possibility is that the mass variation of an object modifies the gravitational energy of the larger thermodynamic system (for example a planet) to which this object belongs.

Coming back to an example evoked in a previous paper ([10]), let us consider the gravitational relations between the Earth and the Moon. Their respective masses (noted $M_1$ and $M_2$), their distant apart (noted R) and the gravitational constant (noted G) have the following values:

$$M_1 = 5.98 \times 10^{24} \text{ kg}$$
$$M_2 = 7.35 \times 10^{22} \text{ kg}$$
$$r = 385\ 000\ 000 \text{ m}$$
$$G = 6.67 \times 10^{-11} \text{ N m}^2 \text{ kg}^{-2}$$

Entering these data into the gravitational equation:

$$E_p = -\frac{GMm}{r} \qquad (62)$$

gives:

$$E_{p\,initial} = -7.614\ 714\ 545 \times 10^{28} \text{ J}$$

This result represents the potential energy of the Earth-Moon system.

Now let us imagine that the average distance Earth-Moon is increased by one meter (385 000 001 m instead of 385 000 000). From eq. 62, we get for the potential energy the new value:

$$E_{p\,final} = -7.614\ 714\ 526 \times 10^{28} \text{ J}$$

The change in potential energy is therefore:

$$\Delta E_p = E_{p\,final} - E_{p\,initial} = 1.978\ 873\ 943 \times 10^{20} \text{ J}$$

This value is positive and, according to the relation $dE = -c^2 dm$, the corresponding change in mass is:

$$\Delta m = -2198.748 \text{ kg}$$

This change in mass concerns the whole Earth-Moon system and can be located inside both bodies or only one. Related to the mass of the Moon, and even more to that of the Earth, such a change appears negligible (respectively 2.99 x $10^{-20}$ and 3,67 x $10^{-22}$). For this reason, eq. 62 gives the same value for Ep whether the term dm is inserted in it or not. The situation is different for the change in distance, because related to the initial value 385 000 000 m, an increase of 1 meter represents a change of 2 x $10^{-9}$. As observed above, this is sufficient for eq. 62 to exhibit a change in potential energy. Nevertheless, if we admit that a change in distance implies a correlative change in mass, we must admit that, reciprocally, a change in mass - even very small - implies a change in distance. Its value can be calculated writing eq. 62 in the form:

$$r = - \frac{GMm}{Ep} \qquad (63)$$

By designating $\Delta r = r_2 - r_1$ the change in distance, we get from eq. 63 the simplified formula:

$$\Delta r = r_2 - r_1 = GMm \left( \frac{1}{Ep_1} - \frac{1}{Ep_2} \right) \qquad (64)$$

If the previously obtained values $Ep_1$ and $Ep_2$ are entered in eq. 64, we get approximately the expected result $\Delta r = 1m$. The relative invariability of the term GMm, compared with the variability of the potential energy Ep and of the distance r is an illustration of the contrast between the concept of "frozen energy", and that of "liberated energy" ([11]).

Although very simple, is seems that this kind of reasoning opens a possibility to extend towards astronomy the link between thermodynamics and relativity suggested by eq. 28.

**2.1. Possible consequences in biology**

It is a matter of fact that the behavior of a living body is not the same after its death as it was before. Knowing that after its death, this behavior is the one corresponding to inert matter, that is to eq. 28, it can be expected that, on the contrary, living matter does not obey eq. 28.

The thermodynamic difference between living matter and inert matter has been studied for a long time by many scientists and has led to the concept of negentropy. Referring to eq. 14, i.e. to the conventional expression of the second law, it consists of the idea that a living system is characterized by a decrease in internal entropy ($dS_i < 0$), instead of the usual increase ($dS_i > 0$) that constitutes the characteristic of inert systems. Introduced in the middle of the XXth century ([12]), this concept is still a subject of active scientific discussion ([13]).

Examined under the light of eq. 28, the problem remains the same, except that the condition $dS_i < 0$ takes the form $dU_{add} < 0$, implying $dm > 0$. Similarly, the condition $dS_i > 0$ takes the form $dU_{add} > 0$, implying $dm < 0$

A few years ago, experiments were performed ([14]) showing a positive change in mass for a closed thermodynamic system made of a mixture of living an inert matter. It is well known, in thermodynamics, that when a closed system is exclusively made of inert matter (a gas contained in a cylinder for example), its exchanges of energy with the surroundings never lead to a detectable change in mass. As a consequence, the observations reported were interesting from a double point of view. The first one because the change in mass was sufficient to be measurable, the second because it was positive.

Curiously, it seems that the results presented by this author have neither been confirmed nor contested. Taking into account the potential incidence of such an information, it would surely be interesting that new experiments be performed.

Referring to eq. 28, a confirmation of the increase in mass would suggest that living matter is able to convert energy to matter. Such a behavior would contrast with that of inert matter, which is supposed here to be characterized by an ability to convert matter to energy.

## - 3 - Conclusions

It is important to note that the hypothesis advanced in this paper is not a rejection of the thermodynamic theory, but an extension and simplification. Both are made possible by the insertion of relativity in the discussion.

The links towards astronomy and biology, briefly evoked above, are examples among others of the kinds of extensions that can be imagined. The idea that every process occurring in nature implies a correlative change in mass is equivalent to say that relativity is omnipresent around us. It is an invitation to search for a close link between thermodynamics and gravitation.

As often felt by students and explicitly mentioned by some authors of textbooks ([15], [16]), the conventional conception of thermodynamics raises conceptual difficulties. Thanks to the simplification allowed by its connection with relativity, it can be expected that the theory would appear more easily accessible to a large scientific readership. Due to the increasing use of the thermodynamic tool in earth sciences, geologists are particularly concerned.

## - 4 - Acknowledgments

I would like to thank the readers who sent me positive comments on my previous papers on the subject. Some of them are among the authors quoted below ([2], [6], [7], (13)]. Although my specialty is geology, rather than physics, I hope that the opening summarized in this paper can be useful.

## - 5 - References

[1]     J-L. Tane, *Thermodynamics and Relativity: A Condensed Explanation of their Close Link*. arxiv.org/pdf/physics/0503106, March 2005

[2]     V. Krasnoholovets and J.-L. Tane. *An extended interpretation of the thermodynamic theory, including an additional energy associated with a decrease in mass*, Int. J. Simulation and Process Modelling 2, Nos. 1/2, 67-79 (2006). arxiv.org/abs/physics/0605094

[3]     R. C. Tolman, *On the Extension of Thermodynamics to General Relativity*
        Proc Natl Acad Sci U S A. 1928 March; 14(3): 268–272.

[4]     R. C. Tolman, *Relativity, Thermodynamics and Cosmology* (1934, 501 pages). Reprinted by Dover Publications (1988)

[5]     Ye Rengui, *The logical connection between special relativity and thermodynamics* Eur. J. Phys. **17** 265-267 (1996)

[6]     P. E. Williams, *Energy and entropy as the fundaments of theoretical Physics*, Entopy 2002, 4 [4], 128-141

[7]     R.C. Gupta, Ruchi Gupta, Sanjay Gupta. *Is Second Law of Thermodynamics Violated for Electron Transition from Lower-Energy Level to Higher-Energy Level* arxiv.org/abs/physics/0310025  (2003)

[8]     C. A. Farías, P. S. Moya, V. A. Pinto. *On the Relationship between Thermodynamics and Special Relativity*. arxiv.org/abs/0712.3793 (2007)

[9]     M. Requardt, *Thermodynamics meets Special Relativity -- or what is real in Physics?* arxiv.org/abs/0801.2639v1 (2008)

[10]    J-L. Tane, *Possible Impact in Astronomy of the Link between Thermodynamics and Relativity,* The General Science Journal, Dec. 2008.

[11]    D. Cassidy, G. Holton and J. Rutherford, *Understanding Physics*, Springer-Verlag, New York, 2002 (comment about "frozen energy" p. 434)

[12]    E. Schrödinger, "*What is Life* ?", New York 1944, Cambridge University Press (reprinted in 1992, with a foreword by Roger Penrose).

[13]    P. Journeau, *New concepts of Dimensions and Consequences*, Frontiers of Fundamental and Computational Physics: 9th International Symposium. AIP Conference Proceedings, Volume 1018, pp. 110-115 (2008*).*

[14]    A. Sorli, *The Additional Mass of Life.* Journal of Theoretics, April/May 2002, Vol.4 No 2.

[15]    D. K. Nordstrom and J. L. Munoz, *Geochemical thermodynamics,* Blackwell Scientific Publications (1986). (See reference to A. Sommerfeld, in the preface).

[16]    G. M. Anderson and D. A. Crerar, *Thermodynamics in geochemistry*, Oxford University Press, (1993). (See reference to H. Reiss in the preface; discussion p 111; reference to R. E. Dickerson, p. 295).